\documentclass[prb,fleqn,twoside,twocolumn,nofootinbib]{revtex4}
\usepackage{graphicx}
\usepackage{amsmath}
\usepackage{amssymb}
\usepackage{amstext}

\begin{document}

\title{Particle-hole pair states of layered materials.}

\author{Lyubov E. Lokot}

\affiliation{Institute of Semiconductor Physics, NAS of Ukraine, 41, Nauky Ave., Kyiv 03028, Ukraine, e-mail: llokot@gmail.com, tel: +380509030899}

\begin{abstract}

In the paper a theoretical study the both the quantized energies of excitonic states and their wave functions in gapped graphene and in monolayer of $\textrm{Mo}\textrm{S}_{2}$ is presented. An integral two-dimensional Schr\"{o}dinger equation of the electron-hole pairing for a particles with electron-hole symmetry of reflection is analytically solved. The solutions of Schr\"{o}dinger equation in momentum space in gapped graphene and in the direct band monolayer of $\textrm{Mo}\textrm{S}_{2}$ by projection the two-dimensional space of momentum on the three-dimensional sphere are found. We analytically solve an integral two-dimensional Schr\"{o}dinger equation of the electron-hole pairing for particles with electron-hole symmetry of reflection and with strong spin-orbit coupling. In monolayer of $\textrm{Mo}\textrm{S}_{2}$ as well as in single-layer graphene (SLG) the electron-hole pairing leads to the exciton insulator states. Calculating an integral two-dimensional Schr\"{o}dinger equation of the electron-hole pairing for bilayer graphene, an exciton insulator states with a gap 3 meV are predicted. The particle-hole symmetry of Dirac equation of layered materials allows perfect pairing between electron Fermi sphere and hole Fermi sphere in the valence band and conduction band and hence driving the Cooper instability.

PACS number(s): 81.05.ue, 81.05.U-, 71.30.+h, 71.10.-w.

\end{abstract}

\maketitle

\section{Introduction}

The graphene and graphene-like systems as well as the $\textrm{M}\textrm{X}_{2}$ (M=Mo, W, X=S, Se) ~\cite{{Novoselov1},{Novoselov2},{Vasko},{Cheiwchanchamnangij},{Komsa},{Zhu},{Peelaers},{Cheiwchanchamnangij1},{Ramasubramaniam}} present a new state of matter of layered materials. The energy bands for graphite was found using "tight-binding" approximation by P.R. Wallace ~\cite{{Wallace}}. In the low-energy limit the single-particle spectrum is Dirac cone similarly to the light cone in relativistic mechanics, where the light speed is replaced by the Fermi velocity $v_{F}$.

In the paper we present a theoretical investigation of excitonic states as well as their wave functions in gapped graphene and in a direct band $\textrm{Mo}\textrm{S}_{2}$. An integral form of the two-dimensional Schr\"{o}dinger equation of Kepler problem in momentum space is solved exactly by projection the two-dimensional space of momentum on the three-dimensional sphere in the paper ~\cite{{Parfitt}}.

The integral Schr\"{o}dinger equation was analytically solved by the projection the three-dimensional momentum space onto the surface of a four-dimensional unit sphere by Fock in 1935 ~\cite{{Fock}}.

We consider the pairing between oppositely charged particles with complex dispersion. The Coulomb interaction leads to the electron-hole bound states scrutiny study of which acquire significant attention in the explanations of superconductivity.

If the exciton binding energy is greater than the flat band gap in narrow-gap semiconductor or semimetal then at sufficiently low temperature the insulator ground state is instable with respect to the exciton formation ~\cite{{Stroucken},{Jerome}}. And excitons may be spontaneously created. In a system undergo a phase transition into a exciton insulator phase similarly to Bardeen-Cooper-Schrieffer (BCS) superconductor. In a single-layer graphene (SLG) and in a single-layer $\textrm{Mo}\textrm{S}_{2}$ the electron-hole pairing leads to the exciton insulator states ~\cite{{LyubovLokot}}.

In the paper an integral two-dimensional Schr\"{o}dinger equation of the electron-hole pairing for particles with complex dispersion is analytically solved. A complex dispersions lead to fundamental difference in exciton insulator states and their wave functions.

A crossing direct-gap like dispersion of single layer of graphene and single layer of $\textrm{Mo}\textrm{S}_{2}$ does not lead to the fundamental differences in the many-particle effects in comparison with w\"{u}rtzite semiconductors ~\cite{{Lokot},{Lokot1}}.

We analytically solve an integral two-dimensional Schr\"{o}dinger equation of the electron-hole pairing for particles with electron-hole symmetry of reflection.

For graphene in vacuum the effective fine structure parameter $\alpha_{G}=\frac{e^{2}}{v_{F}\hbar\varepsilon\sqrt{\pi}}=1.23$. For graphene in substrate $\alpha_{G}=0.77$, when the permittivity of graphene in substrate is estimated to be $\varepsilon=1.6$ ~\cite{{Alicea}}. It means the prominent Coulomb effects ~\cite{{Sharapov}}.

It is known that the Coulomb interaction leads to the semimetal-exciton insulator transition, where  gap is opened by electron-electron exchange interaction ~\cite{{Jerome},{Stroucken1},{Kadi},{Malic}}. The perfect host combines a small gap and a large exciton binding energy ~\cite{{Jerome},{Stroucken}}.

In graphene as well as in $\textrm{Mo}\textrm{S}_{2}$ the existing of bound pair states are still subject matter of researches ~\cite{{Gamayun},{Gamayun1},{Berman},{Berman1},{Hartmann}}.

It is known ~\cite{{Min}} in the weak-coupling limit ~\cite{{Sak}}, exciton condensation is a consequence of the Cooper instability of materials with electron-hole symmetry of reflection inside identical Fermi surface. The identical Fermi surfaces is a consequence of the particle-hole symmetry of Dirac equation. The room temperature superfluidity are shown to be calculated for bilayer graphene ~\cite{{Stroucken},{Min}}.

The particle-hole symmetry of Dirac equation allows perfect pairing between electron Fermi sphere and hole Fermi sphere in the opposite layer and hence driving the Cooper instability. In the weak-coupling limit in graphene with the occupied conduction-band states and empty valence-band states inside identical Fermi surfaces in band structure, the exciton condensation is a consequence of the Cooper instability.

\section{Theoretical study}

\subsection{Graphene}

In the honeycomb lattice of graphene with two carbon atoms per unit cell the space group is $D_{3h}^{1}$ ~\cite{{Malard}}:

\begin{widetext}
\begin{tabular}{cccccccc} \hline\hline
\multicolumn{1}{c}{$D_{3h}^{1}$} &
\multicolumn{1}{c}{${\{E|0\}}$} &
\multicolumn{1}{c}{${\{C^{(+,-)}_{3}|0\}}$} &
\multicolumn{1}{c}{${\{C_{2}'^{(A,B,C)}|0\}}$} &
\multicolumn{1}{c}{${\{\sigma_{h}|\tau\}}$} &
\multicolumn{1}{c}{${\{S^{(-,+)}_{3}|\tau\}}$} &
\multicolumn{1}{c}{${\{\sigma^{(A,B,C)}_{v}|\tau\}}$} &
\multicolumn{1}{c}{$ $} \\
\hline
$K_{3}^{+}$ & 2 & -1 & 0 & 2 & -1 & 0 & $ $ \\
$g^{2}$ & ${\{E|0\}}$ & ${\{C^{(+,-)}_{3}|0\}}$ & ${\{E|0\}}$ & ${\{E|0\}}$ & ${\{S^{(-,+)}_{3}|\tau\}}$ & ${\{E|0\}}$ & $ $ \\
$\chi^{2}(g)$ & 4 & 1 & 0 & 4 & 1 & 0 & $K^{+}_{1}+K^{+}_{2}+K^{+}_{3}$\\
$\chi(g^{2})$ & 2 & -1 & 2 & 2 & -1 & 2 & $ $\\
$\frac{1}{2}[\chi^{2}(g)+\chi(g^{2})]$ & 3 & 0 & 1 & 3 & 0 & 1 & $K^{+}_{1}+K^{+}_{3}$\\
$\frac{1}{2}\{\chi^{2}(g)-\chi(g^{2})\}$ & 1 & 1 & -1 & 1 & 1 & -1 & $K^{+}_{2}$\\ \hline\hline
\end{tabular}
\end{widetext}

The direct production of two irreducible presentations of wave function and wave vector of difference $\kappa-K$ or $\kappa-K'$ expansion is $K^{+}_{3}\times\,K_{3}^{+\star}$ and can be expanded on

%Fig1
\begin{figure}%[h,p!]
\includegraphics*[bb=5 10 1000 600,width=4in]{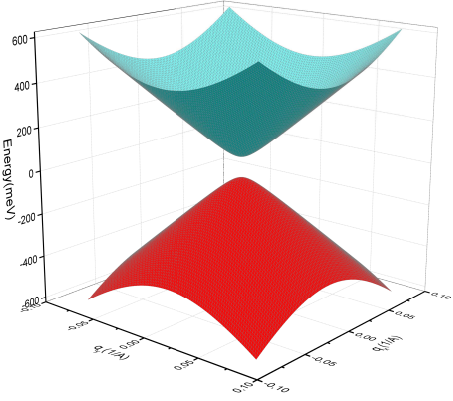}
\caption{Single-particle spectrum of gapped graphene}
\end{figure}

\begin{equation}
p^{\alpha}: \tau_{\psi}\times\,\tau_{k}=(K^{+}_{1}+K^{+}_{2}+K^{+}_{3})\times\,K^{+}_{3}=K^{+}_{3}\times\,K^{+}_{3}.
\end{equation}

In the low-energy limit the single-particle spectrum is Dirac cone. The Hamiltonian of graphene ~\cite{{Wallace}}

\begin{equation}
\hat{H}=\frac{\Delta}{2}\hat{\sigma_{z}}+v_{F}(\tau\,q_{x}\hat{\sigma_{x}}+q_{y}\hat{\sigma_{y}}),
\end{equation}
where
$\Delta$ is band gap of graphene, $q_{x}$, $q_{y}$ are Cartesian components of a wave vector, $\tau=\pm\,1$ is the valley index, $v_{F}=1\times10^{6}$ m/s is the graphene Fermi velocity, $\hat{\sigma_{x}}$, $\hat{\sigma_{y}}$, $\hat{\sigma_{z}}$ are Pauli matrices (here we assume that $\hbar=1$).

The dispersion of energy bands may be found in the form ~\cite{{Wallace}}
\begin{equation}
\begin{array}{c}
\epsilon_{\pm}=\pm\,\frac{\Delta}{2}\sqrt{1+\frac{4v_{F}^{2}q^{2}}{\Delta^{2}}},\\
\end{array}
\end{equation}
where
$q=\sqrt{q_{x}^{2}+q_{y}^{2}}$.

The Schr\"{o}dinger equation for the calculating of exciton states can be written in the general form
\begin{equation}
(\epsilon(\textbf{q}^{2})+q_{0}^{2})\Phi(\textbf{q})=\frac{1}{\pi}\int{\frac{\Phi(\textbf{q}')}{|\textbf{q}-\textbf{q}'|}}d\textbf{q}',
\end{equation}
where $q_{0}^{2}=-\epsilon$, $\epsilon$ is a quantized energy. We look for the bound states and hence the energy will be negative.

An integral form of the two-dimensional Schr\"{o}dinger equation in momentum space for the gapped graphene is solved exactly by projection the two-dimensional space of momentum on the three-dimensional sphere.

For the gapped single layer graphene
\begin{equation}
\begin{array}{c}
\frac{\epsilon(q^{2})+q_{0}^{2}}{q^{2}+q_{0}^{2}}=\pm\frac{\Delta}{4q_{0}^{2}}\sqrt{(1-\cos{\theta})^{2}+\frac{4v_{F}^{2}}{\Delta^{2}}\,q_{0}^{2}(\sin{\theta})^{2}}+\\
+\frac{1-\cos{\theta}}{2},
\end{array}
\end{equation}
where an each point on sphere is defined of two spherical angles $\theta$, $\phi$, which are knitted with a momentum $\textbf{q}$ ~\cite{{Fock},{Parfitt}}. A space angle $\Omega$ may be found as surface element on sphere $d\Omega=\sin({\theta})d\theta\,d\phi=(\frac{2q_{0}}{q^{2}+q_{0}^{2}})^{2}d\textbf{q}$ ~\cite{{Fock},{Parfitt}}.
A spherical angle $\theta$ and a momentum $\textbf{q}$ are shown ~\cite{{Fock},{Parfitt}} to be knitted as
\begin{equation}
\cos{\theta}=\frac{q^{2}-q_{0}^{2}}{q^{2}+q_{0}^{2}},\,\sin{\theta}=\frac{2qq_{0}}{q^{2}+q_{0}^{2}},\,q^{2}=q_{0}^{2}(\frac{1+\cos{\theta}}{1-\cos{\theta}}).
\end{equation}

Using spherical symmetry the solution of integral Schr\"{o}dinger equation can look for in the form
\begin{equation}
\Phi(\textbf{q})=\sqrt{q_{0}}(\frac{2q_{0}}{q^{2}+q_{0}^{2}})^{3/2}\sum_{l=0}^{\infty}A_{l}Y_{l}^{0}(\theta,\phi),
\end{equation}
where
\begin{equation}
\begin{array}{c}
Y_{l}^{0}(\theta,\phi)=\sqrt{\frac{2l+1}{4\pi}}P_{l}^{0}(\cos{\theta}).
\end{array}
\end{equation}

Since ~\cite{{Parfitt}}
\begin{widetext}
\begin{equation}
\begin{array}{c}
\frac{(q^{2}+q_{0}^{2})^{1/2}(q'^{2}+q_{0}^{2})^{1/2}}{2q_{0}}\frac{1}{|\textbf{q}-\textbf{q}'|}=\sum_{\lambda=0}^{\infty}\sum_{\mu=-\lambda}^{\lambda}\frac{4\pi}{2\lambda+1}Y_{\lambda}^{\mu}(\theta,\phi)Y_{\lambda}^{\mu,\ast}(\theta',\phi'),
\end{array}
\end{equation}
then substituting (7), (9) in (4), can find equation
\begin{equation}
\begin{array}{c}
\frac{\epsilon(q^{2})+q_{0}^{2}}{q^{2}+q_{0}^{2}}\sum_{l=0}^{\infty}A_{l}Y_{l}^{0}(\theta,\phi)=\frac{2}{q_{0}}\sum_{l=0}^{\infty}\sum_{\lambda=0}^{\infty}\sum_{\mu=-\lambda}^{\lambda}\int\frac{1}{2\lambda+1}Y_{\lambda}^{\mu}(\theta,\phi)Y_{\lambda}^{\mu,\ast}(\theta',\phi')Y_{l}^{0}(\theta',\phi')A_{l}(\frac{2q_{0}}{q'^{2}+q_{0}^{2}})^{2}d\textbf{q}'.
\end{array}
\end{equation}

The integral equations for gapped SLG based on Eq. (5) may be found in the form
\begin{equation}
\begin{array}{c}
\int(\mp\frac{\Delta}{4q_{0}^{2}}\sqrt{(1-\cos{\theta})^{2}+\frac{4v_{F}^{2}}{\Delta^{2}}\,q_{0}^{2}(\sin{\theta})^{2}}+\frac{1-\cos{\theta}}{2})\sum_{l=0}^{\infty}A_{l}Y_{l}^{0}(\theta,\phi)Y_{k}^{n,\ast}(\theta,\phi)d\Omega=\\
=\frac{2}{q_{0}}\int\sum_{\lambda=0}^{\infty}\sum_{\mu=-\lambda}^{\lambda}\sum_{l'=0}^{\infty}\frac{1}{2\lambda+1}Y_{\lambda}^{\mu}(\theta,\phi)Y_{\lambda}^{\mu,\ast}(\theta',\phi')Y_{l'}^{0}(\theta',\phi')Y_{k}^{n,\ast}(\theta,\phi)d\Omega\,d\Omega'A_{l'}.
\end{array}
\end{equation}

Since ~\cite{{Fock1}}
\begin{equation}
\begin{array}{c}
\cos{\theta}P_{l}^{m}(\cos{\theta})=\frac{\sqrt{l^{2}-m^{2}}}{\sqrt{4l^{2}-1}}P_{l-1}^{m}(\cos{\theta})+\frac{\sqrt{(l+1)^{2}-m^{2}}}{\sqrt{4(l+1)^{2}-1}}P_{l+1}^{m}(\cos{\theta}),
\end{array}
\end{equation}
\begin{equation}
\begin{array}{c}
\sin{\theta}P_{l}^{m}(\cos{\theta})=\frac{\sqrt{(l-m)(l-m-1)}}{\sqrt{4l^{2}-1}}P_{l-1}^{m+1}(\cos{\theta})+\frac{\sqrt{(l+m+1)(l+m+2)}}{\sqrt{4(l+1)^{2}}-1}P_{l+1}^{m+1}(\cos{\theta}),
\end{array}
\end{equation}
\end{widetext}
then solutions of the integral equation (10) for the energies and wave functions correspondingly can be found analytically with taken into account the normalization condition $(\frac{1}{2\pi})^2\int{\frac{q^{2}+q_{0}^{2}}{2q_{0}^{2}}|\Phi(\textbf{q})|^{2}d\textbf{q}}=1$.

From equation (11) one can obtain the eigenvalue and eigenfunction problem and using a condition $\frac{4v_{F}^{2}}{\Delta^{2}}\,q_{0}^{2}>>1$
one can find recurrence relation
\begin{equation}
\frac{1}{2}(l+\frac{1}{2})A_{l}+\frac{1}{q_{0}}A_{l}+\frac{1}{2}A_{l-1}(l+\frac{1}{2})a_{l}+\frac{1}{2}A_{l+1}(l+\frac{1}{2})b_{l}=0.
\end{equation}

The solutions of the quantized series in excitonic Rydbergs where Ry=$m_{r}e^{4}/(\epsilon^{2}\hbar^{2})=34.72$ meV, $m_{r}$ is the reduced mass of an electron-hole pair, and wave functions of the integral equation (11) one can find in the form
\begin{equation}
\epsilon_{0}=-\frac{1}{(\frac{1}{4}+\frac{1}{2}(1+\frac{1}{2})a_{1})^{2}},
\end{equation}
\begin{equation}
\epsilon_{1}=-\frac{1}{(\frac{1}{2}(1+\frac{1}{2})+\frac{1}{4}b_{0}+\frac{1}{2}(2+\frac{1}{2})a_{2})^{2}},
\end{equation}
\begin{equation}
\epsilon_{2}=-\frac{1}{(\frac{1}{2}(2+\frac{1}{2})+\frac{1}{2}(1+\frac{1}{2})b_{1}+\frac{1}{2}(3+\frac{1}{2})a_{3})^{2}},
\end{equation}
\begin{equation}
\epsilon_{3}=-\frac{1}{(\frac{1}{2}(3+\frac{1}{2})+\frac{1}{2}(2+\frac{1}{2})b_{2}+\frac{1}{2}(4+\frac{1}{2})a_{4})^{2}},
\end{equation}
\begin{equation}
\begin{array}{c}
\Phi_{l}(\cos{\theta})=\sqrt{\frac{2\pi}{(q_{0l})^3}}\sum_{n=0}^{\infty}(1-\cos{\theta})^{3/2}P_{n}^{0}(\cos{\theta}),
\end{array}
\end{equation}
where $q_{0l}^{2}=-\epsilon_{l}$, $l=0,1,2,3,4,....$,
\begin{equation}
a_{l}=\frac{1}{2\pi}\sqrt{\frac{2(l-1)+1}{2}}\sqrt{\frac{2}{2l+1}}\frac{l}{\sqrt{4l^{2}+1}},
\end{equation}
\begin{equation}
b_{l}=\frac{1}{4\pi}\sqrt{2(l+1)+1}\sqrt{2l+1}\frac{l+1}{\sqrt{4(l+1)^{2}-1}}.
\end{equation}

%Fig2
\begin{figure}%[h,p!]
\includegraphics*[bb=5 10 1000 600,width=4in]{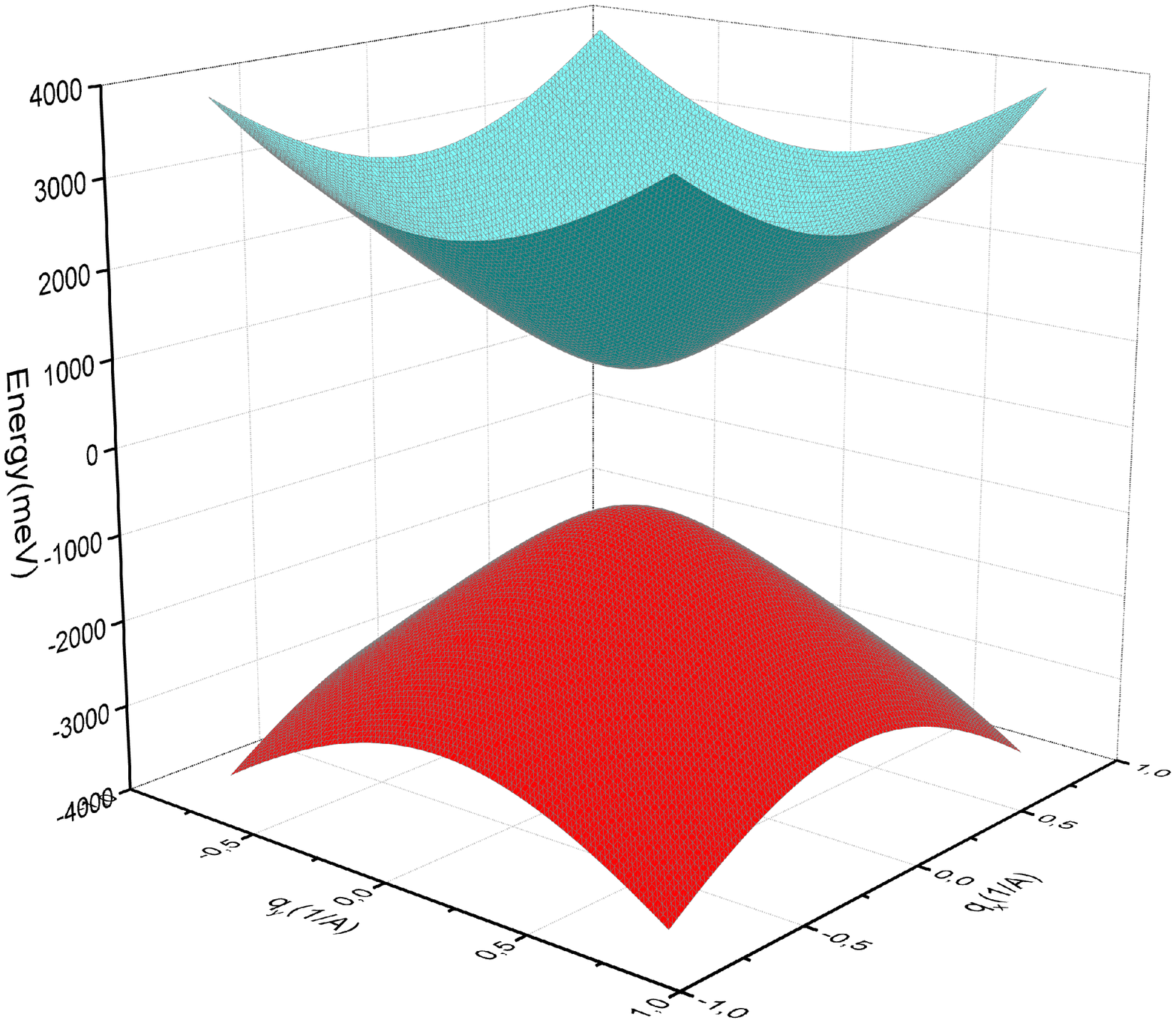}
\caption{Single-particle spectrum of the two-degenerated conduction band and spin-orbit splitting upper valence band of $\textrm{Mo}\textrm{S}_{2}$}
\end{figure}

Table 1. Quantized spectral series of the excitonic states in meV, band gap of graphene in meV, effective reduced mass of electron-hole pair, the effective fine structure parameter, excitonic Rydberg in meV.

\begin{tabular}{cccccccc} \hline\hline
\multicolumn{1}{c}{$\epsilon_{0}$} &
\multicolumn{1}{c}{$\epsilon_{1}$} &
\multicolumn{1}{c}{$\epsilon_{2}$} &
\multicolumn{1}{c}{$\epsilon_{3}$} &
\multicolumn{1}{c}{$\Delta$} &
\multicolumn{1}{c}{$m_{r}$} &
\multicolumn{1}{c}{$\alpha_{G}$} &
\multicolumn{1}{c}{$Ry$} \\
\hline
440.22 & 48.66 & 15.73 & 7.14 & 100 & 0.0033 & 0.77 & 34.72 \\ \hline\hline
\end{tabular}

\subsection{$\textrm{Mo}\textrm{S}_{2}$}

In the honeycomb lattice of $\textrm{Mo}\textrm{S}_{2}$ the space group is $C_{3h}^{1}$ ~\cite{{Xiao}}:

\begin{widetext}
\begin{tabular}{cccccccc} \hline\hline
\multicolumn{1}{c}{$C^{1}_{3h}$} &
\multicolumn{1}{c}{${\{E|0\}}$} &
\multicolumn{1}{c}{${\{C_{3}|0\}}$} &
\multicolumn{1}{c}{${\{C^{2}_{3}|0\}}$} &
\multicolumn{1}{c}{${\{\sigma_{h}|\tau\}}$} &
\multicolumn{1}{c}{${\{S_{3}|\tau\}}$} &
\multicolumn{1}{c}{${\{S^{5}_{3}|\tau\}}$} &
\multicolumn{1}{c}{$ $} \\
\hline
$ $ & 2 & -1 & -1 & 2 & -1 & -1 & $(B_{1}^{+}+B_{2}^{+})$ \\
$g^{2}$ & ${\{E|0\}}$ & ${\{C^{2}_{3}|0\}}$ & ${\{C_{3}|0\}}$ & ${\{E|0\}}$ & ${\{S^{5}_{3}|\tau\}}$ & ${\{S_{3}|\tau\}}$ & $ $ \\
$\chi_{g}^{2}(g)$ & 4 & 1 & 1 & 4 & 1 & 1 & $ $\\
$\chi(g^{2})$ & 2 & -1 & -1 & 2 & -1 & -1  & $ $\\
$\frac{1}{2}[\chi^{2}(g)+\chi(g^{2})]$ & 3 & 0 & 0 & 3 & 0 & 0 & $(A^{+}+B_{1}^{+}+B_{2}^{+})$\\
$\frac{1}{2}[\chi^{2}(g)-\chi(g^{2})]$ & 1 & 1 & 1 & 1 & 1 & 1 & $(A^{+})$\\ \hline\hline
\end{tabular}
\end{widetext}

The direct production of two irreducible presentations of wave function and wave vector of difference $\kappa-K$ or $\kappa-K'$ expansion with taken into account time inversion can be expanded on

\begin{equation}
\begin{array}{c}
p^{\alpha}: \tau_{v}\times\,\tau_{\psi}=(B_{1}^{+}+B_{2}^{+})\times\,(A^{+}+B_{1}^{+}+B_{2}^{+})=\\
=B_{1}^{+}\times\,B_{1}^{+}+B_{2}^{+}\times\,B_{2}^{+},
\end{array}
\end{equation}

\begin{equation}
\{p^{\alpha}p^{\beta}\}: (A^{+})\times\,(A^{+}+B_{1}^{+}+B_{2}^{+})=A^{+}\times\,A^{+}.
\end{equation}

%Fig3
\begin{figure}%[h,p!]
\includegraphics*[bb=5 10 1000 600,width=4in]{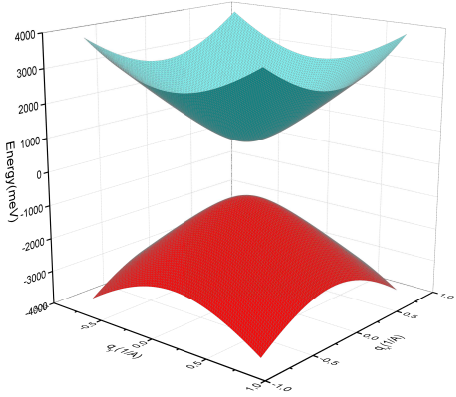}
\caption{Single-particle spectrum of the two-degenerated conduction band and spin-orbit splitting lower valence band of $\textrm{Mo}\textrm{S}_{2}$}
\end{figure}

The Hamiltonian of $\textrm{Mo}\textrm{S}_{2}$ ~\cite{{Xiao}}:
\begin{equation}
\hat{H}=at(\tau\,q_{x}\hat{\sigma_{x}}+q_{y}\hat{\sigma_{y}})+\frac{\Delta}{2}\hat{\sigma_{z}}-\nu\,\tau\frac{\hat{\sigma_{z}}-1}{2}\hat{s_{z}},
\end{equation}
where $\Delta$ the energy gap, $\tau=\pm\,1$ is the valley index, $a$ is the lattice constant, $t$ the effective hopping integral, $2\nu$ is the spin splitting at the valence band top caused by the spin-orbit coupling (SOC), $s_{z}$ is the Pauli matrix for spin.

The dispersion of bands may be found in the form:
\begin{equation}
\begin{array}{c}
\epsilon_{\pm}=\pm\,\frac{\Delta}{2}\sqrt{1+\frac{4a^{2}t^{2}q^{2}}{\Delta^{2}}},\\
\end{array}
\end{equation}

\begin{equation}
\begin{array}{c}
\epsilon_{\pm}=-\frac{\nu}{4}\pm\,\frac{\sqrt{\frac{\nu^{2}}{4}+\Delta\,\nu+\Delta^{2}}}{2}\sqrt{1+\frac{4a^{2}t^{2}q^{2}}{\frac{\nu^{2}}{4}+\Delta\,\nu+\Delta^{2}}}.\\
\end{array}
\end{equation}

For the two-degenerated conduction band and spin-orbit splitting upper valence band one can find
\begin{equation}
\begin{array}{c}
\frac{\epsilon(q^{2})+q_{0}^{2}}{q^{2}+q_{0}^{2}}=\pm\frac{\Delta}{4q_{0}^{2}}\sqrt{(1-\cos{\theta})^{2}+\frac{4a^{2}t^{2}}{\Delta^{2}}\,q_{0}^{2}(\sin{\theta})^{2}}+\frac{1-\cos{\theta}}{2}.
\end{array}
\end{equation}

For the two-degenerated conduction band and spin-orbit splitting lower valence band one can find
\begin{widetext}
\begin{equation}
\begin{array}{c}
\frac{\epsilon(q^{2})+q_{0}^{2}}{q^{2}+q_{0}^{2}}=\pm\frac{\nu}{8q_{0}^{2}}\sqrt{(1-\cos{\theta})^{2}+\frac{4a^{2}t^{2}}{\frac{\nu^{2}}{4}+\Delta\,\nu+\Delta^{2}}\,q_{0}^{2}(\sin{\theta})^{2}}+\frac{1-\cos{\theta}}{2}(1-\frac{\nu}{4q_{0}^{2}}).
\end{array}
\end{equation}

The integral equation for the two-degenerated conduction band and spin-orbit splitting upper valence band may be found correspondingly in the form substituting (27) in (10)

\begin{equation}
\begin{array}{c}
\int(\mp\frac{\Delta}{4q_{0}^{2}}\sqrt{(1-\cos{\theta})^{2}+\frac{4a^{2}t^{2}}{\Delta^{2}}\,q_{0}^{2}(\sin{\theta})^{2}}+\frac{1-\cos{\theta}}{2})\sum_{l=0}^{\infty}A_{l}Y_{l}^{0}(\theta,\phi)Y_{k}^{n,\ast}(\theta,\phi)d\Omega=\\
=\frac{2}{q_{0}}\int\sum_{\lambda=0}^{\infty}\sum_{\mu=-\lambda}^{\lambda}\sum_{l'=0}^{\infty}\frac{1}{2\lambda+1}Y_{\lambda}^{\mu}(\theta,\phi)Y_{\lambda}^{\mu,\ast}(\theta',\phi')Y_{l'}^{0}(\theta',\phi')Y_{k}^{n,\ast}(\theta,\phi)d\Omega\,d\Omega'A_{l'}.
\end{array}
\end{equation}

From equation (29) one can obtain the eigenvalue and eigenfunction problem and using a condition $\frac{4a^{2}t^{2}}{\Delta^{2}}\,q_{0}^{2}>>1$
one can find
\begin{equation}
\frac{1}{2}(l+\frac{1}{2})A_{l}+\frac{1}{q_{0}}A_{l}+\frac{1}{2}A_{l-1}(l+\frac{1}{2})a_{l}+\frac{1}{2}A_{l+1}(l+\frac{1}{2})b_{l}=0.
\end{equation}
\end{widetext}

%Fig4
\begin{figure}%[h,p!]
\includegraphics*[bb=5 10 1000 600,width=4in]{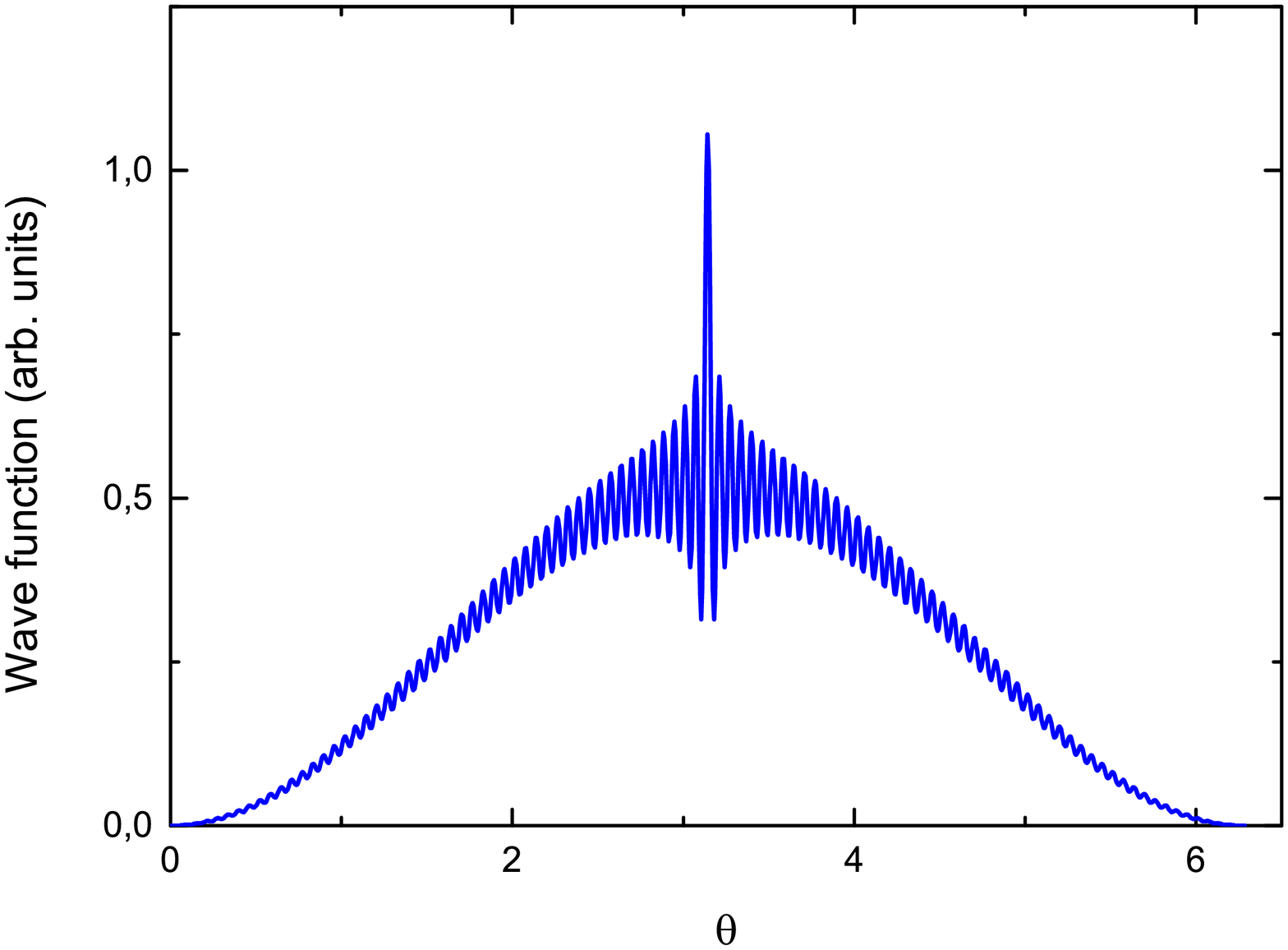}
\caption{Wave function for graphene and single-particle spectrum of the two-degenerated conduction band and spin-orbit splitting upper valence band of $\textrm{Mo}\textrm{S}_{2}$ for quantum number $l=0$}
\end{figure}

%Fig5
\begin{figure}%[h,p!]
\includegraphics*[bb=5 10 1000 600,width=4in]{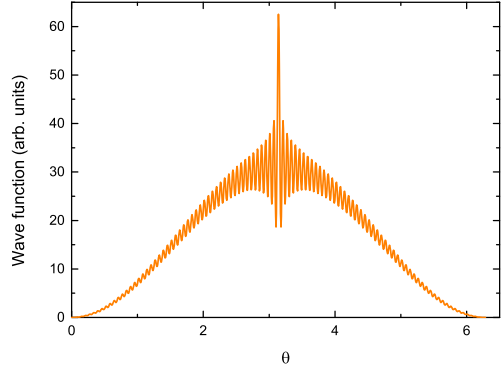}
\caption{Wave function for single-particle spectrum of the two-degenerated conduction band and spin-orbit splitting lower valence band of $\textrm{Mo}\textrm{S}_{2}$ for quantum number $l=0$}
\end{figure}

The solutions of the quantized spectral series in excitonic Rydbergs where Ry=$m_{r}e^{4}/(\epsilon^{2}\hbar^{2})=342.16$ meV, $m_{r}$ is the reduced mass of an electron-hole pair, and wave functions of the integral equation (29) one can find in the form

\begin{equation}
\epsilon_{0}=-\frac{1}{(\frac{1}{4}+\frac{1}{2}(1+\frac{1}{2})a_{1})^{2}},
\end{equation}

\begin{equation}
\epsilon_{1}=-\frac{1}{(\frac{1}{2}(1+\frac{1}{2})+\frac{1}{4}b_{0}+\frac{1}{2}(2+\frac{1}{2})a_{2})^{2}},
\end{equation}

\begin{equation}
\epsilon_{2}=-\frac{1}{(\frac{1}{2}(2+\frac{1}{2})+\frac{1}{2}(1+\frac{1}{2})b_{1}+\frac{1}{2}(3+\frac{1}{2})a_{3})^{2}},
\end{equation}

\begin{equation}
\epsilon_{3}=-\frac{1}{(\frac{1}{2}(3+\frac{1}{2})+\frac{1}{2}(2+\frac{1}{2})b_{2}+\frac{1}{2}(4+\frac{1}{2})a_{4})^{2}},
\end{equation}

\begin{equation}
\begin{array}{c}
\Phi_{l}(\cos{\theta})=\sqrt{\frac{2\pi}{(q_{0l})^3}}\sum_{n=0}^{\infty}(1-\cos{\theta})^{3/2}P_{n}^{0}(\cos{\theta}).
\end{array}
\end{equation}

Table 2. Quantized spectral series of the excitonic states in meV for the two-degenerated conduction band and spin-orbit splitting upper valence band, effective reduced mass of electron-hole pair, excitonic Rydberg in meV.

\begin{tabular}{ccccccc} \hline\hline
\multicolumn{1}{c}{$\epsilon_{0}$} &
\multicolumn{1}{c}{$\epsilon_{1}$} &
\multicolumn{1}{c}{$\epsilon_{2}$} &
\multicolumn{1}{c}{$\epsilon_{3}$} &
\multicolumn{1}{c}{$m_{r}$} &
\multicolumn{1}{c}{$Ry$} \\
\hline
4338.78 & 479.60 & 155.06 & 70.37 & 0.146 & 342.16 \\ \hline\hline
\end{tabular}

\begin{widetext}
The integral equation for the two-degenerated conduction band and spin-orbit splitting lower valence band may be found correspondingly in the form substituting (28) in (10)

\begin{equation}
\begin{array}{c}
\int(\mp\frac{\nu}{8q_{0}^{2}}\sqrt{(1-\cos{\theta})^{2}+\frac{a^{2}t^{2}}{\frac{\nu^{2}}{4}+\Delta\,\nu+\Delta^{2}}\,q_{0}^{2}(\sin{\theta})^{2}}+\frac{1-\cos{\theta}}{2}(1-\frac{\nu}{4q_{0}^{2}}))\sum_{l=0}^{\infty}A_{l}Y_{l}^{0}(\theta,\phi)Y_{k}^{n,\ast}(\theta,\phi)d\Omega=\\
=\frac{2}{q_{0}}\int\sum_{\lambda=0}^{\infty}\sum_{\mu=-\lambda}^{\lambda}\sum_{l'=0}^{\infty}\frac{1}{2\lambda+1}Y_{\lambda}^{\mu}(\theta,\phi)Y_{\lambda}^{\mu,\ast}(\theta',\phi')Y_{l'}^{0}(\theta',\phi')Y_{k}^{n,\ast}(\theta,\phi)d\Omega\,d\Omega'A_{l'}.
\end{array}
\end{equation}

From equation (36) one can obtain the eigenvalue and eigenfunction problem and using a condition $\frac{4a^{2}t^{2}}{\Delta^{2}}\,q_{0}^{2}>>1$
one can find

\begin{equation}
\frac{1}{2}(l+\frac{1}{2})A_{l}+\frac{1}{q_{0}}A_{l}+\frac{1}{2}A_{l-1}(l+\frac{1}{2})a_{l}(1-\frac{\nu}{4q_{0}^{2}})+\frac{1}{2}A_{l+1}(l+\frac{1}{2})b_{l}(1-\frac{\nu}{4q_{0}^{2}})=0.
\end{equation}
\end{widetext}

The solutions of the quantized energies and wave functions of the integral equation one can find in the form
\begin{equation}
\epsilon_{0\pm}=-\frac{9a_{1}^{2}\nu^{2}}{64(-1\pm\sqrt{1+\frac{3}{16}a_{1}\nu(3a_{1}+1)})^{2}},
\end{equation}
\begin{equation}
\epsilon_{1\pm}=-\frac{\nu^{2}(5a_{2}+b_{0})^{2}}{64(-1\pm\sqrt{1+\frac{\nu}{16}(3+(5a_{2}+b_{0}))(5a_{2}+b_{0})})^{2}},
\end{equation}
\begin{equation}
\epsilon_{2\pm}=-\frac{\nu^{2}(7a_{3}+3b_{1})^{2}}{64(-1\pm\sqrt{1+\frac{\nu}{16}(5+(7a_{3}+3b_{1}))(7a_{3}+3b_{1})})^{2}},
\end{equation}
\begin{equation}
\epsilon_{3\pm}=-\frac{\nu^{2}(8a_{4}+5b_{2})^{2}}{64(-1\pm\sqrt{1+\frac{\nu}{16}(7+(8a_{4}+5b_{2}))(8a_{4}+5b_{2})})^{2}},
\end{equation}
\begin{equation}
\begin{array}{c}
\Phi_{l}(\cos{\theta})=\sqrt{\frac{2\pi}{(q_{0l})^3}}\sum_{n=0}^{\infty}(1-\cos{\theta})^{3/2}P_{n}^{0}(\cos{\theta}).
\end{array}
\end{equation}

Table 3. Quantized spectral series of the excitonic states in meV for the two-degenerated conduction band and spin-orbit splitting lower valence band.

\begin{tabular}{ccccccc} \hline\hline
\multicolumn{1}{c}{$\epsilon_{0}$} &
\multicolumn{1}{c}{$\epsilon_{1}$} &
\multicolumn{1}{c}{$\epsilon_{2}$} &
\multicolumn{1}{c}{$\epsilon_{3}$} \\
\hline
4305.39 & 446.29 & 123.45 & 41.32 \\ \hline\hline
\end{tabular}

\subsection{Bilayer graphene}

In the bilayer graphene the space group is $D_{3}^{1}$, the point group of $K$ point is $D_{3}$ ~\cite{{Malard}}:

\begin{widetext}
\begin{tabular}{ccccc} \hline\hline
\multicolumn{1}{c}{$D_{3}^{1}$} &
\multicolumn{1}{c}{${\{E|0\}}$} &
\multicolumn{1}{c}{${\{C^{(+,-)}_{3}|0\}}$} &
\multicolumn{1}{c}{${\{C_{2}'^{(A,B,C)}|0\}}$} &
\multicolumn{1}{c}{$ $} \\
\hline
$\tau_{v}$ & 2 & -1 & 0 & $\Gamma_{3}$ \\
$g^{2}$ & ${\{E|0\}}$ & ${\{C^{(+,-)}_{3}|0\}}$ & ${\{E|0\}}$ & $ $ \\
$\chi^{2}(g)$ & 4 & 1 & 0 & $ $\\
$\chi(g^{2})$ & 2 & -1 & 2 & $ $\\
$\frac{1}{2}[\chi^{2}(g)+\chi(g^{2})]$ & 3 & 0 & 1 & $\Gamma_{1}+\Gamma_{3}$\\
$\frac{1}{2}\{\chi^{2}(g)-\chi(g^{2})\}$ & 1 & 1 & -1 & $\Gamma_{2}$\\ \hline\hline
\end{tabular}

\begin{tabular}{cccccccc} \hline\hline
\multicolumn{1}{c}{$D_{3}^{1}$} &
\multicolumn{1}{c}{${\{E|0\}}$} &
\multicolumn{1}{c}{${\{C^{(+,-)}_{3}|0\}}$} &
\multicolumn{1}{c}{${\{C_{2}'^{(A,B,C)}|0\}}$} &
\multicolumn{1}{c}{$Q{\{E|0\}}$} &
\multicolumn{1}{c}{$Q{\{C^{(+,-)}_{3}|0\}}$} &
\multicolumn{1}{c}{$Q{\{C_{2}'^{(A,B,C)}|0\}}$} &
\multicolumn{1}{c}{$ $} \\
\hline
$\tau_{\psi}$ & 2 & 1 & 0 & -2 & -1 & 0 & $\Gamma_{4}$ \\
$g^{2}$ & ${\{E|0\}}$ & ${\{C^{(+,-)}_{3}|0\}}$ & ${\{E|0\}}$ & ${\{E|0\}}$ & ${\{C^{(+,-)}_{3}|0\}}$ & ${\{E|0\}}$ & $ $ \\
$\chi^{2}(g)$ & 4 & 1 & 0 & 4 & 1 & 0 & $ $\\
$\chi(g^{2})$ & 2 & -1 & -2 & 2 & -1 & -2 & $ $\\
$\frac{1}{2}[\chi^{2}(g)+\chi(g^{2})]$ & 3 & 0 & -1 & 3 & 0 & -1 & $\Gamma_{2}+\Gamma_{3}$\\
$\frac{1}{2}\{\chi^{2}(g)-\chi(g^{2})\}$ & 1 & 1 & 1 & 1 & 1 & 1 & $\Gamma_{1}$\\ \hline\hline
\end{tabular}
\end{widetext}

The direct production of two irreducible presentations of wave function and wave vector of difference $\kappa-K$ or $\kappa-K'$ expansion for the wave vector including time inversion can be expanded on

%Fig6
\begin{figure}%[h,p!]
\includegraphics*[bb=5 10 1000 600,width=4in]{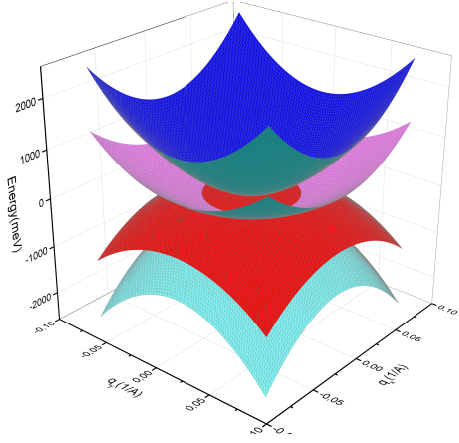}
\caption{Single-particle spectrum of bilayer graphene}
\end{figure}

\begin{equation}
p^{\alpha}: \tau_{\psi}\times\,\tau_{q}=(\Gamma_{1}+\Gamma_{3})\times\,(\Gamma_{2}+\Gamma_{3})=\Gamma_{3}\times\,\Gamma_{3},
\end{equation}

for the square of wave vector
\begin{equation}
p^{\alpha}p^{\beta}: \tau_{\psi}\times\,\tau_{q^{2}}=(\Gamma_{1}+\Gamma_{3})\times\,(\Gamma_{1})=\Gamma_{1}\times\,\Gamma_{1}.
\end{equation}

The Hamiltonian of bilayer graphene

\begin{equation}
\hat{H}=\frac{\Delta}{2}\hat{\sigma_{z}}+\varsigma\,Aq^{2}\hat{I}+v_{F}(\tau\,q_{x}\hat{\sigma_{x}}+q_{y}\hat{\sigma_{y}}),
\end{equation}
where
$\Delta$ is band gap of graphene, $\varsigma=\pm$ is the particle index, $m=1/A$ is effective mass, $\hat{I}$ is unit matrix, $q_{x}$, $q_{y}$ are Cartesian components of a wave vector, $\tau=\pm\,1$ is the valley index, $v_{F}=1\times10^{6}$ m/s is the graphene Fermi velocity, $\hat{\sigma_{x}}$, $\hat{\sigma_{y}}$, $\hat{\sigma_{z}}$ are Pauli matrices (here we assume that $\hbar=1$).

The dispersion of energy bands may be found in the form
\begin{equation}
\begin{array}{c}
\epsilon_{\pm}=\pm(Aq^{2}\pm\,\frac{\Delta}{2}\sqrt{1+\frac{4v_{F}^{2}q^{2}}{\Delta^{2}}}),\\
\end{array}
\end{equation}
where
$q=\sqrt{q_{x}^{2}+q_{y}^{2}}$.

The Schr\"{o}dinger equation for the calculating of exciton states can be written in the general form
\begin{equation}
(\epsilon(\textbf{q}^{2})+q_{0}^{2})\Phi(\textbf{q})=\frac{1}{\pi}\int{\frac{\Phi(\textbf{q}')}{|\textbf{q}-\textbf{q}'|}}d\textbf{q}',
\end{equation}
where $q_{0}^{2}=-\epsilon$, $\epsilon$ is a quantized energy. We look for the bound states and hence the energy will be negative.

An integral form of the two-dimensional Schr\"{o}dinger equation in momentum space for the gapped graphene is solved exactly by projection the two-dimensional space of momentum on the three-dimensional sphere.

When an each point on sphere is defined of two spherical angles $\theta$, $\phi$, which are knitted with a momentum $\textbf{q}$ ~\cite{{Fock},{Parfitt}}. A space angle $\Omega$ may be found as surface element on sphere $d\Omega=\sin({\theta})d\theta\,d\phi=(\frac{2q_{0}}{q^{2}+q_{0}^{2}})^{2}d\textbf{q}$ ~\cite{{Fock},{Parfitt}}.
A spherical angle $\theta$ and a momentum $\textbf{q}$ are shown ~\cite{{Fock},{Parfitt}} to be knitted as
\begin{equation}
\cos{\theta}=\frac{q^{2}-q_{0}^{2}}{q^{2}+q_{0}^{2}},\,\sin{\theta}=\frac{2qq_{0}}{q^{2}+q_{0}^{2}},\,q^{2}=q_{0}^{2}(\frac{1+\cos{\theta}}{1-\cos{\theta}}).
\end{equation}
For the bilayer graphene
\begin{equation}
\begin{array}{c}
\frac{\epsilon(q^{2})+q_{0}^{2}}{q^{2}+q_{0}^{2}}=\pm\frac{\Delta}{4q_{0}^{2}}\sqrt{(1-\cos{\theta})^{2}+\frac{4v_{F}^{2}}{\Delta^{2}}\,q_{0}^{2}(\sin{\theta})^{2}}\pm\\
\pm\,A+(1\mp\,A)(\frac{1-\cos(\theta)}{2}).
\end{array}
\end{equation}

Using spherical symmetry the solution of integral Schr\"{o}dinger equation can look for in the form
\begin{equation}
\Phi(\textbf{q})=\sqrt{q_{0}}(\frac{2q_{0}}{q^{2}+q_{0}^{2}})^{3/2}\sum_{l=0}^{\infty}A_{l}Y_{l}^{0}(\theta,\phi),
\end{equation}
where
\begin{equation}
\begin{array}{c}
Y_{l}^{0}(\theta,\phi)=\sqrt{\frac{2l+1}{4\pi}}P_{l}^{0}(\cos{\theta}).
\end{array}
\end{equation}

Since ~\cite{{Parfitt}}
\begin{widetext}
\begin{equation}
\begin{array}{c}
\frac{(q^{2}+q_{0}^{2})^{1/2}(q'^{2}+q_{0}^{2})^{1/2}}{2q_{0}}\frac{1}{|\textbf{q}-\textbf{q}'|}=\sum_{\lambda=0}^{\infty}\sum_{\mu=-\lambda}^{\lambda}\frac{4\pi}{2\lambda+1}Y_{\lambda}^{\mu}(\theta,\phi)Y_{\lambda}^{\mu,\ast}(\theta',\phi'),
\end{array}
\end{equation}
then substituting (50), (52) in (47), can find equation
\begin{equation}
\begin{array}{c}
\frac{\epsilon(q^{2})+q_{0}^{2}}{q^{2}+q_{0}^{2}}\sum_{l=0}^{\infty}A_{l}Y_{l}^{0}(\theta,\phi)=\frac{2}{q_{0}}\sum_{l=0}^{\infty}\sum_{\lambda=0}^{\infty}\sum_{\mu=-\lambda}^{\lambda}\int\frac{1}{2\lambda+1}Y_{\lambda}^{\mu}(\theta,\phi)Y_{\lambda}^{\mu,\ast}(\theta',\phi')Y_{l}^{0}(\theta',\phi')A_{l}(\frac{2q_{0}}{q'^{2}+q_{0}^{2}})^{2}d\textbf{q}'.
\end{array}
\end{equation}

The integral equations for bilayer graphene based on Eq. (49) may be found in the form
\begin{equation}
\begin{array}{c}
\int(\pm\frac{\Delta}{4q_{0}^{2}}\sqrt{(1-\cos{\theta})^{2}+\frac{4v_{F}^{2}}{\Delta^{2}}\,q_{0}^{2}(\sin{\theta})^{2}}\pm\,A+(1\mp\,A)(\frac{1-\cos(\theta)}{2}))\sum_{l=0}^{\infty}A_{l}Y_{l}^{0}(\theta,\phi)Y_{k}^{n,\ast}(\theta,\phi)d\Omega=\\
=\frac{2}{q_{0}}\int\sum_{\lambda=0}^{\infty}\sum_{\mu=-\lambda}^{\lambda}\sum_{l'=0}^{\infty}\frac{1}{2\lambda+1}Y_{\lambda}^{\mu}(\theta,\phi)Y_{\lambda}^{\mu,\ast}(\theta',\phi')Y_{l'}^{0}(\theta',\phi')Y_{k}^{n,\ast}(\theta,\phi)d\Omega\,d\Omega'A_{l'}.
\end{array}
\end{equation}

Since ~\cite{{Fock1}}
\begin{equation}
\begin{array}{c}
\cos{\theta}P_{l}^{m}(\cos{\theta})=\frac{\sqrt{l^{2}-m^{2}}}{\sqrt{4l^{2}-1}}P_{l-1}^{m}(\cos{\theta})+\frac{\sqrt{(l+1)^{2}-m^{2}}}{\sqrt{4(l+1)^{2}-1}}P_{l+1}^{m}(\cos{\theta}),
\end{array}
\end{equation}
\begin{equation}
\begin{array}{c}
\sin{\theta}P_{l}^{m}(\cos{\theta})=\frac{\sqrt{(l-m)(l-m-1)}}{\sqrt{4l^{2}-1}}P_{l-1}^{m+1}(\cos{\theta})+\frac{\sqrt{(l+m+1)(l+m+2)}}{\sqrt{4(l+1)^{2}}-1}P_{l+1}^{m+1}(\cos{\theta}),
\end{array}
\end{equation}
\end{widetext}
then solutions of the integral equation (54) for the energies and wave functions correspondingly can be found analytically with taken into account the normalization condition $(\frac{1}{2\pi})^2\int{\frac{q^{2}+q_{0}^{2}}{2q_{0}^{2}}|\Phi(\textbf{q})|^{2}d\textbf{q}}=1$.

From equation (54) one can obtain the eigenvalue and eigenfunction problem and using a condition $\frac{4v_{F}^{2}}{\Delta^{2}}\,q_{0}^{2}>>1$
one can find recurrence relation

\begin{equation}
\begin{array}{c}
\frac{1\pm\,A}{2}(l+\frac{1}{2})A_{l}+\frac{1}{q_{0}}A_{l}+\\
+A_{l-1}\frac{(1\mp\,A)}{2}(l+\frac{1}{2})a_{l}+A_{l+1}\frac{(1\mp\,A)}{2}(l+\frac{1}{2})b_{l}=0.
\end{array}
\end{equation}

%Fig7
\begin{figure}%[h,p!]
\includegraphics*[bb=5 10 1000 600,width=4in]{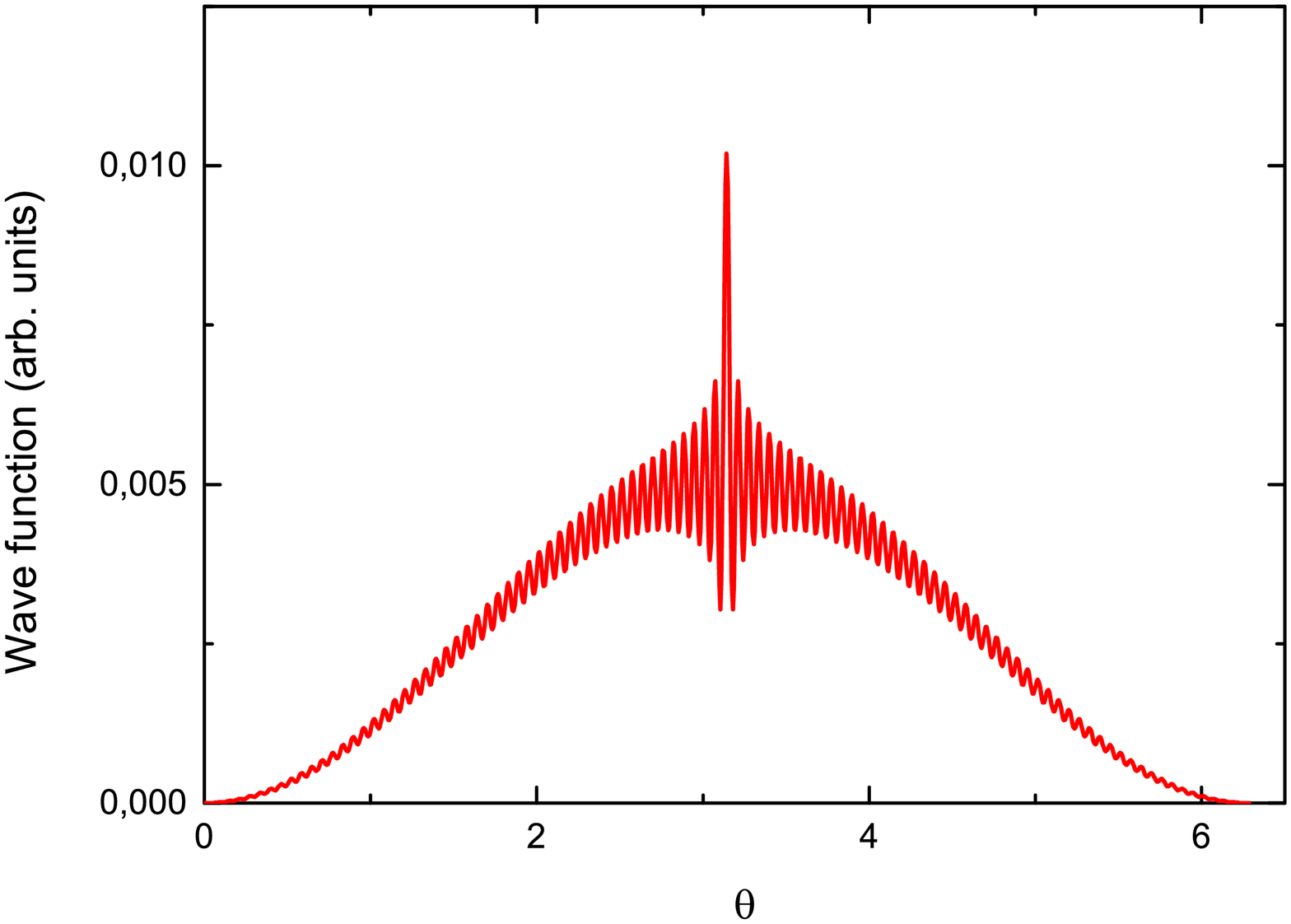}
\caption{Wave function for bilayer graphene for quantum number $l=0$ for quantized spectral energy $\epsilon_{0+}$}
\end{figure}

%Fig8
\begin{figure}%[h,p!]
\includegraphics*[bb=5 10 1000 600,width=4in]{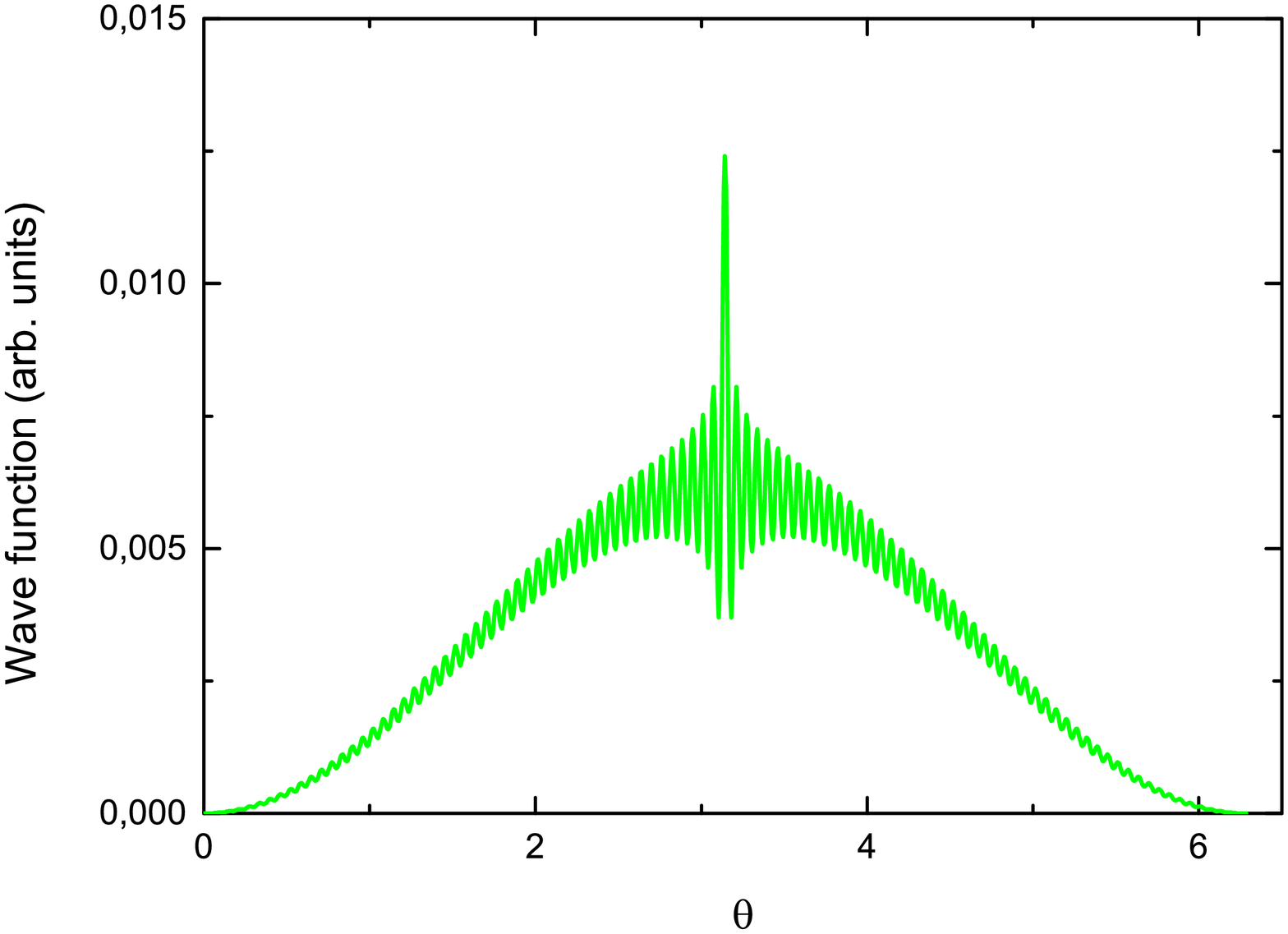}
\caption{Wave function for bilayer graphene for quantum number $l=0$ for quantized spectral energy $\epsilon_{0-}$}
\end{figure}

The solutions of the quantized series in excitonic Rydbergs and wave functions of the integral equation (54) one can find in the form
\begin{equation}
\epsilon_{0\pm}=-\frac{1}{(\frac{(1\pm\,A)}{4}+\frac{(1\mp\,A)}{2}(1+\frac{1}{2})a_{1})^{2}},
\end{equation}
\begin{equation}
\epsilon_{1\pm}=-\frac{1}{(\frac{(1\pm\,A)}{2}(1+\frac{1}{2})+\frac{(1\mp\,A)}{4}b_{0}+\frac{(1\mp\,A)}{2}(2+\frac{1}{2})a_{2})^{2}},
\end{equation}
\begin{equation}
\begin{array}{c}
\epsilon_{2\pm}=\\
=-\frac{1}{(\frac{(1\pm\,A)}{2}(2+\frac{1}{2})+\frac{(1\mp\,A)}{2}(1+\frac{1}{2})b_{1}+\frac{(1\mp\,A)}{2}(3+\frac{1}{2})a_{3})^{2}},
\end{array}
\end{equation}
\begin{equation}
\begin{array}{c}
\epsilon_{3\pm}=\\
=-\frac{1}{(\frac{(1\pm\,A)}{2}(3+\frac{1}{2})+\frac{(1\mp\,A)}{2}(2+\frac{1}{2})b_{2}+\frac{(1\mp\,A)}{2}(4+\frac{1}{2})a_{4})^{2}},
\end{array}
\end{equation}
\begin{equation}
\begin{array}{c}
\Phi_{l}(\cos{\theta})=\sqrt{\frac{2\pi}{(q_{0l})^3}}\sum_{n=0}^{\infty}(1-\cos{\theta})^{3/2}P_{n}^{0}(\cos{\theta}),
\end{array}
\end{equation}
where $q_{0l}^{2}=-\epsilon_{l}$, $l=0,1,2,3,4,....$, excitonic Rydberg Ry=$m_{r}e^{4}/(\epsilon^{2}\hbar^{2})=346.61$ meV, $m_{r}$ is the reduced mass of an electron-hole pair, $m=\frac{\gamma_{1}}{2v_{F}^{2}}$ ~\cite{Stroucken} is the electron (hole) mass,

Table 4. Quantized spectral series of the excitonic states in meV, band gap of graphene in meV, effective reduced mass of electron-hole pair, excitonic Rydberg in meV.

\begin{tabular}{cccccccc} \hline\hline
\multicolumn{1}{c}{$\epsilon_{0}$} &
\multicolumn{1}{c}{$\epsilon_{1}$} &
\multicolumn{1}{c}{$\epsilon_{2}$} &
\multicolumn{1}{c}{$\epsilon_{3}$} &
\multicolumn{1}{c}{$\Delta$} &
\multicolumn{1}{c}{$m_{r}$} &
\multicolumn{1}{c}{$Ry$} \\
\hline
4.41 & 0.49 & 0.20 & 0.12 & 3 & 0.0255 & 346.61 \\ \hline\hline
\end{tabular}

Table 5. Quantized spectral series of the excitonic states in meV.

\begin{tabular}{ccccc} \hline\hline
\multicolumn{1}{c}{$\epsilon_{0}$} &
\multicolumn{1}{c}{$\epsilon_{1}$} &
\multicolumn{1}{c}{$\epsilon_{2}$} &
\multicolumn{1}{c}{$\epsilon_{3}$} & \\
\hline
5.03 & 0.56 & 0.24 & 0.15  \\ \hline\hline
\end{tabular}

\section{Results and discussions}

The integral Schr\"{o}dinger equation for a parabolic bands was analytically solved by the projection the three-dimensional momentum space onto the surface of a four-dimensional unit sphere by Fock in 1935 ~\cite{{Fock}}.

In the paper an integral two-dimensional Schr\"{o}dinger equation of the electron-hole pairing for particles with complex dispersion is analytically solved. A complex dispersion leads to fundamental difference in the energy of exciton insulator states and their wave functions.

A crossing direct-gap like dispersion of single layer of graphene as well as in single layer of $\textrm{Mo}\textrm{S}_{2}$ does not lead to the fundamental differences in the many-particle effects in comparison with w\"{u}rtzite semiconductors ~\cite{{Lokot},{Lokot1}}.

We analytically solve an integral two-dimensional Schr\"{o}dinger equation of the electron-hole pairing for particles with electron-hole symmetry of reflection.

It is known that the Coulomb interaction leads to the semimetal-exciton insulator transition, where  gap is opened by electron-electron exchange interaction ~\cite{{Jerome},{Stroucken1},{Kadi},{Malic}}. The perfect host combines a small gap and a large exciton binding energy ~\cite{{Jerome},{Stroucken}}.

We consider the pairing between oppositely charged particles in gapped graphene. The Coulomb interaction leads to the electron-hole bound states scrutiny study of which acquire significant attention in the explanations of superconductivity.

It is known ~\cite{{Stroucken},{Jerome}} if the exciton binding energy is greater than the flat band gap in narrow-gap semiconductor or semimetal then at sufficiently low temperature the insulator ground state is instable concerning to the exciton formation with follow up spontaneous production of excitons. In a system undergo a phase transition into a exciton insulator phase similarly to BCS superconductor. In a SLG as well as in a single-layer $\textrm{Mo}\textrm{S}_{2}$ and in bilayer graphene the electron-hole pairing leads to the exciton insulator states.

The particle-hole symmetry of Dirac equation of layered materials allows perfect pairing between electron Fermi sphere and hole Fermi sphere in the valence band and conduction band and hence driving the Cooper instability. In the weak-coupling limit in graphene with the occupied conduction-band states and empty valence-band states inside identical Fermi surfaces in band structure, the exciton condensation is a consequence the Cooper instability.

\section{Conclusions}

In this paper we found the solution the integral Schr\"{o}dinger equation in a momentum space of two interacting via a Coulomb potential Dirac particles that form the exciton in gapped graphene, in a single-layer $\textrm{Mo}\textrm{S}_{2}$ and in bilayer graphene.

In low-energy limit this problem is solved analytically. We obtained the energy dispersion and wave function of the exciton in gapped graphene, in monolayer $\textrm{Mo}\textrm{S}_{2}$ and in bilayer graphene. The excitons were considered as a system of two oppositely charge Dirac particles interacting via a Coulomb potential.

We solve this problem in a momentum space because on the whole the center-of-mass and the relative motion of the two Dirac particles can not be separated.

We analytically solve an integral two-dimensional Schr\"{o}dinger equation of the electron-hole pairing for particles with electron-hole symmetry of reflection. An integral form of the two-dimensional Schr\"{o}dinger equation in momentum space for gapped graphene, for monolayer $\textrm{Mo}\textrm{S}_{2}$ and in bilayer graphene is solved exactly by projection the two-dimensional space of momentum on the three-dimensional sphere.

In the SLG as well as in the monolayer $\textrm{Mo}\textrm{S}_{2}$ the electron-hole pairing leads to the exciton insulator states. The exciton insulator states with a gap 3 meV are shown to be found calculating an integral two-dimensional Schr\"{o}dinger equation of the electron-hole pairing for bilayer graphene.

\section{Appendix}

Table 6. The irreducible representational of $D^{1}_{3h}$ ~\cite{Mildred}.
\begin{widetext}
\begin{tabular}{cccccccc} \hline\hline
\multicolumn{1}{c}{$D^{1}_{3h}$} &
\multicolumn{1}{c}{${\{E|0\}}$} &
\multicolumn{1}{c}{${\{C^{(+,-)}_{3}|0\}}$} &
\multicolumn{1}{c}{${\{C_{2}'^{(A,B,C)}|0\}}$} &
\multicolumn{1}{c}{${\{\sigma_{h}|\tau\}}$} &
\multicolumn{1}{c}{${\{S^{(-,+)}_{3}|\tau\}}$} &
\multicolumn{1}{c}{${\{\sigma^{(A,B,C)}_{v}|\tau\}}$} &
\multicolumn{1}{c}{$ $} \\
\hline
$K_{1}^{+}$ & 1 & 1 & 1 & 1 & 1 & 1 & $x^{2}+y^{2},\,z^{2}$ \\
$K_{2}^{+}$ & 1 & 1 & -1 & 1 & 1 & -1 & $J_{z}$ \\
$K_{3}^{+}$ & 2 & -1 & 0 & 2 & -1 & 0 & $(x,\,y)$\\
$K_{1}^{-}$ & 1 & 1 & 1 & -1 & -1 & -1 & $$\\
$K_{2}^{-}$ & 1 & 1 & -1 & -1 & -1 & 1 & $z$\\
$K_{3}^{-}$ & 2 & -1 & 0 & -2 & 1 & 0 & $(x^{2}-y^{2}, xy),\,(J_{x},J_{y})$\\ \hline\hline
\end{tabular}
\end{widetext}

Table 7. The irreducible representational of $C^{1}_{3h}$ ~\cite{Mildred}.
\begin{widetext}
\begin{tabular}{cccccccc} \hline\hline
\multicolumn{1}{c}{$C^{1}_{3h}$} &
\multicolumn{1}{c}{${\{E|0\}}$} &
\multicolumn{1}{c}{${\{C_{3}|0\}}$} &
\multicolumn{1}{c}{${\{C^{2}_{3}|0\}}$} &
\multicolumn{1}{c}{${\{\sigma_{h}|\tau\}}$} &
\multicolumn{1}{c}{${\{S_{3}|\tau\}}$} &
\multicolumn{1}{c}{${\{S^{5}_{3}|\tau\}}$} &
\multicolumn{1}{c}{$ $} \\
\hline
$A^{+}$ & 1 & 1 & 1 & 1 & 1 & 1 & $J_{z},\,x^{2}+y^{2},\,z^{2}$ \\
$A^{-}$ & 1 & 1 & 1 & -1 & -1 & -1 & $z$ \\
$B_{1}^{+}$ & 1 & $\varepsilon$ & $\varepsilon^{2}$ & 1 & $\varepsilon$ & $\varepsilon^{2}$ & $x+iy$\\
$B_{1}^{-}$ & 1 & $\varepsilon$ & $\varepsilon^{2}$ & -1 & $-\varepsilon$ & $-\varepsilon^{2}$ & $J_{x}+iJ{y}$\\
$B_{2}^{+}$ & 1 & $\varepsilon^{2}$ & $\varepsilon$ & 1 & $\varepsilon^{2}$ & $\varepsilon$ & $x-iy$\\
$B_{2}^{-}$ & 1 & $\varepsilon^{2}$ & $\varepsilon$ & -1 & $-\varepsilon^{2}$ & $-\varepsilon$ & $J_{x}-iJ{y}$\\ \hline\hline
\end{tabular}
\end{widetext}

$\varepsilon=\exp(2\pi\,i/3)$

Table 8. The irreducible representational of $D^{1}_{3}$ ~\cite{Mildred}.
\begin{widetext}
\begin{tabular}{cccccccc} \hline\hline
\multicolumn{1}{c}{$D^{1}_{3}$} &
\multicolumn{1}{c}{${\{E|0\}}$} &
\multicolumn{1}{c}{${\{C^{(+,-)}_{3}|0\}}$} &
\multicolumn{1}{c}{${\{C_{2}'^{(A,B,C)}|0\}}$} &
\multicolumn{1}{c}{$ $} \\
\hline
$\Gamma_{1}$ & 1 & 1 & 1 & $x^{2}+y^{2},\,z^{2}$ \\
$\Gamma_{2}$ & 1 & 1 & -1 & $J_{z},\,z$ \\
$\Gamma_{3}$ & 2 & -1 & 0 & $(xz,\,yz)\,(x^{2}-y^{2},\,xy),\,(x,\,y),\,(J_{x},\,J_{y})$\\
\hline\hline
\end{tabular}
\end{widetext}

\section{Creation of bielectron of Dirac cone: the tachyon solution in magnetic field.}

Schr\"{o}dinger equation for pair of two massless Dirac particles when magnetic field is applied in Landau gauge is solved exactly. In this case the separation of center of mass and relative motion is obtained. Landau quantization $\epsilon=\pm\,B\sqrt{l}$ for pair of two Majorana fermions coupled via a Coulomb potential from massless chiral Dirac equation in cylindric coordinate is found. The root ambiguity in energy spectrum leads into Landau quantization for bielectron, when the states in which the one simultaneously exists are allowed. The tachyon solution with imaginary energy in Cooper problem ($\epsilon^{2}<0$) is found. The continuum symmetry of Dirac equation allows perfect pairing between electron Fermi spheres when magnetic field is applied in Landau gauge creating a Cooper pair.

\section{Introduction}

The graphene ~\cite{{Novoselov},{Vasko},{Zhang}} presents a new state of matter of layered materials. The energy bands for graphite was found using "tight-binding" approximation by P.R. Wallace ~\cite{{Wallace}}. In the low-energy limit the single-particle spectrum is Dirac cone similarly to the light cone in relativistic physics, where the light velocity is substituted by the Fermi velocity $v_{F}$ and describes by the massless Dirac equation.

The graphene is the single graphite layer, i. e. two-dimensional graphite plane of thickness of single atom. The graphene lattice resembles a honeycomb lattice. The graphene lattice one can consider like into the composite of two triangular sublattices. In 1947 Wallace in "tight-binding" approximation consider a graphite which consist off the graphene blocks with taken into account the overlap only the nearest $\pi$-electrons.

The two-dimensional nature of graphene and the space and point symmetries of graphene acquire of the reason for the massless electron motion since lead into massless Dirac equation (Majorana fermions) ~\cite{{Wallace},{Semenoff}}. At low-energy limit the single particle spectrum forms with $\pi$-electron carbon orbital and consist off completely occupation valence cone and completely empty conduction cone, which have cone like shape with single Dirac point. In Dirac point the existing an electron as well as a hole is proved. The state in Dirac cone is double degenerate with taken into account a spin.

The existing of the massless Dirac fermions in graphene was proved based on the unconventional quantum Hall effect. The reason of creation the integer Hall conductivity ~\cite{{Zheng},{Gusynin1},{Gusynin2},{Peres}} is derived from Berry phase ~\cite{{Berry},{Carmier}}.

When the magnetic field is applied perpendicularly into graphene plane the lowest (n=0) Landau level has the energy $\pm\Delta$ in two nonequivalent cones $K_{\mp}$, correspondingly ~\cite{Gusynin3}. In the paper ~\cite{Gusynin3} the Dirac mass via a splitting value is found when Zeeman coupling is absence. These properties of the lowest Landau level which distribute between particles and antiparticles in equal parts are base of the integer quantum Hall effect in graphene ~\cite{Gusynin3}. For $n\geq1$ an all Landau levels are fourfold degenerate. For $n=0$ a states in both cones are twofold degenerate with energies $\pm\Delta$ with taken into account a spin ~\cite{Gusynin3}.

\begin{widetext}

\section{Solution of massless chiral Dirac equation for pair of two Majorana fermions coupled via a Coulomb potential in magnetic field in Landau gauge.}

Calculation model of the graphene reflects continuum symmetry of QED$_{2+1}$ including Lorentz group. SU(2) symmetry are shown to be found similar to chiral in the paper ~\cite{Gusynin3}. Hence is conserving quantum number of chirality.

The energy bands for graphene was found using "tight-binding" approximation in the papers ~\cite{{Wallace},{Gusynin3}}.

Calculate the quantized Landau energy as well as the wave function of the Majorana particles in cylindrical coordinate in magnetic field in Landau gauge. Enter the production and annihilation operators as following:
\begin{equation}\label{deq0}
\begin{array}{cccc}
\hat{c}^{\dag}=\sqrt{B}(-\frac{\partial}{\partial\,\xi_{2}}+\xi_{1}),\\
\hat{c}=\sqrt{B}(\frac{\partial}{\partial\,\xi_{1}}+\xi_{2}),\\
\end{array}
\end{equation}
where $\xi_{1}=\frac{\sqrt{B}}{2}\zeta$, $\xi_{2}=\frac{\sqrt{B}}{2}\eta$, $\zeta=x+iy$, $\eta=x-iy$, which satisfies the commutator relation:
\begin{equation}
[\hat{c}^{\dag},\hat{c}]=2B.
\end{equation}

Hence for noninteracting Dirac particles we write the massless Dirac equation in the form:

\begin{equation}
\frac{i}{\sqrt{2}}\hbar\,v_{F}\left\|
\begin{array}{cc}
0 & \hat{c}^{\dag}\\
\hat{c} & 0 \\
\end{array}
\right\|\left\|
\begin{array}{cc}
\Psi_{1}\\
\Psi_{2}\\
\end{array}
\right\|=\varepsilon\left\|
\begin{array}{cc}
\Psi_{1}\\
\Psi_{2}\\
\end{array}
\right\|.\end{equation}

The Schr\"{o}dinger equation for the reduced energy can be rewritten in the form:

\begin{equation}
\frac{1}{2}\hat{c}^\dag\,\hat{c}\,\Psi=\epsilon^{2}\Psi.
\end{equation}

For graphene in vacuum the effective fine structure parameter $\alpha_{G}=\frac{e^{2}}{v_{F}\hbar\kappa\sqrt{\pi}}=1.23$. For graphene in substrate $\alpha_{G}=0.77$, when the permittivity of graphene in substrate is estimated to be $\kappa=1.6$ ~\cite{{Alicea}}. It means the prominent Coulomb effects. Hence the Coulomb potential may be found in the form ~\cite{Novikov}:
\begin{equation}\label{deq01}
V(\rho)=\hbar\,v_{F}\frac{\alpha}{\rho},
\end{equation}

where $v_{F}=10^{6}$ m/s is the graphene Fermi velocity (here we assume that $\hbar=1$).

%Fig1
\begin{figure}
\includegraphics*[bb=5 10 1000 600,width=5in]{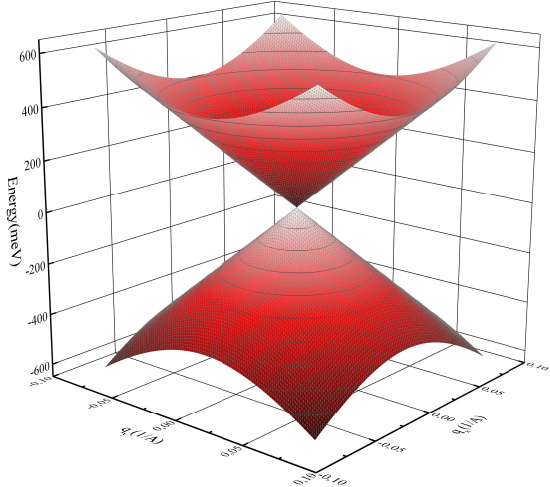}
\caption{(Color online) Single-particle spectrum of graphene for massless Dirac fermions (Majorana fermions).}
\end{figure}

When magnetic field is applied perpendicularly into graphene plane in z axis along field distribution. The vector potential in the gauge ~\cite{Landau} $\textbf{A}=\frac{1}{2}[\textbf{Hr}]$ has a components $A_{\varphi}=H\rho/2$, $A_{\rho}=A_{z}=0$ and Schr\"{o}dinger equation:
\begin{equation}\label{deq1}
\begin{array}{cccc}
-4\frac{\partial^{2}\Psi}{\partial\,\rho^{2}}-\frac{4}{\rho}\frac{\partial\Psi}{\partial\rho}-\frac{4}{\rho^{2}}\frac{\partial^{2}\Psi}{\partial\varphi^{2}}-\frac{4i}{\rho^{2}}\frac{\partial\Psi}{\partial\varphi}
-2iB\frac{\partial\Psi}{\partial\varphi}+\frac{B^{2}}{4}\rho^{2}\Psi-2B\Psi=2\epsilon^{2}\Psi.
\end{array}
\end{equation}

The Schr\"{o}dinger equation \eqref{deq1} with including the Coulomb potential Eq. \eqref{deq01} one can rewritten in the form:

\begin{equation}\label{deq*}
\begin{array}{cccc}
-4\frac{\partial^{2}\Psi}{\partial\,\rho^{2}}-\frac{4}{\rho}\frac{\partial\Psi}{\partial\rho}-\frac{4}{\rho^{2}}\frac{\partial^{2}\Psi}{\partial\varphi^{2}}-\frac{4i}{\rho^{2}}\frac{\partial\Psi}{\partial\varphi}
-2iB\frac{\partial\Psi}{\partial\varphi}+\frac{B^{2}}{4}\rho^{2}\Psi-2B\Psi+\frac{\alpha^{2}}{\rho^{2}}\Psi=2\epsilon^{2}\Psi.
\end{array}
\end{equation}

The solution Eq. \eqref{deq*} with including the Coulomb potential Eq. \eqref{deq01} can look for in the form:
\begin{equation}
\Psi_{n,k}=\frac{1}{\sqrt{2\pi}}R(\rho)e^{im\varphi}.
\end{equation}

Substituting the solution in Eq. \eqref{deq*}, one can find for the radial function the following equation
\begin{equation}
R''+\frac{1}{\rho}R'+(\beta-\frac{k^{2}}{\rho^{2}}-2\gamma\,m-\gamma^{2}\rho^{2})R=0,
\end{equation}
where $\beta=(\epsilon^{2}+B)/2$, $\gamma=B/4$, $k^2=m^2+m-\alpha^2$.
Entering the new independent variable $\xi=\gamma\rho^2$, the equation \eqref{deq*} can be rewritten in the form:
\begin{equation}\label{deq2}
\xi\,R''+R'+(\frac{\xi}{4}+\lambda-\frac{k^{2}}{4\xi})R=0,
\end{equation}
where $\lambda=\frac{\beta}{4\gamma}-\frac{m}{2}$.
At $\xi\rightarrow\infty$ conduct of sought for function are shown to be found as following $e^{-\xi/2}$, and at $\xi\rightarrow0$ like $\xi^{k/2}$.

The solution of the Eq. \eqref{deq2} can look for in the form:
\begin{equation}\label{deq3}
R=e^{-\xi/2}\xi^{k/2}\varpi(\xi).
\end{equation}

To substitute the solution \eqref{deq3} in the Eq. \eqref{deq2} it is necessarily to find as follows
\begin{equation}\label{deq4}
R''=\varpi''e^{-\xi/2}\xi^{k/2}+2\varpi'\frac{\partial}{\partial\,\xi}e^{-\xi/2}\xi^{k/2}+\varpi\frac{\partial^{2}}{\partial\,\xi^{2}}e^{-\xi/2}\xi^{k/2},
\end{equation}
\begin{equation}\label{deq5}
R'=\varpi'e^{-\xi/2}\xi^{k/2}+\varpi\frac{\partial}{\partial\,\xi}e^{-\xi/2}\xi^{k/2}.
\end{equation}
Since
\begin{equation}\label{deq6}
\frac{\partial}{\partial\,\xi}e^{-\xi/2}\xi^{k/2}=-\frac{1}{2}e^{-\xi/2}\xi^{k/2}+\frac{k}{2}e^{-\xi/2}\xi^{k/2-1},
\end{equation}
\begin{equation}\label{deq7}
\begin{array}{cccc}
\frac{\partial^{2}}{\partial\,\xi^{2}}e^{-\xi/2}\xi^{k/2}=\frac{1}{4}e^{-\xi/2}\xi^{k/2}-\frac{k}{2}e^{-\xi/2}\xi^{k/2-1}+\frac{k}{2}(\frac{k}{2}-1)e^{-\xi/2}\xi^{k/2-2}.\\
\end{array}
\end{equation}
Substituting \eqref{deq4}, \eqref{deq5}, \eqref{deq6}, \eqref{deq7} into the Eq. \eqref{deq2} we find the equation for $\varpi(\xi)$ as follows
\begin{equation}\label{deq8}
\xi\varpi''+\varpi'(1+k-\xi)+\varpi(\lambda-\frac{k+1}{2})=0.
\end{equation}

Hence for $\varpi(\xi)$ we derive the equation for confluent hypergeometric function:
\begin{equation}
\varpi=F(-(\lambda-\frac{k+1}{2}),k+1,\xi).
\end{equation}
From the condition of finite of the wave function one can find the energy spectrum in the form:
\begin{equation}
\epsilon=\pm\,B\sqrt{l},
\end{equation}
where $l=2n+m+k$, $\lambda-\frac{k+1}{2}=n$,  $m=0,1,2,3,...$. The wave function expressed via the associated Laguerre polynomial:
\begin{equation}
R_{n,k}=[\frac{B(n-k)!}{2n!^{3}(n-k+1)}]^{1/2}\psi_{n,k},
\end{equation}
where
\begin{equation}
\psi_{n,k}=e^{-\frac{\xi}{2}}\xi^{\frac{k}{2}}L_{n}^{k}(\xi).
\end{equation}

Because the solution for the wave functions for the pair of two massless Dirac particles when magnetic field is applied in Landau gauge one can express via the product of the two identical wave functions one can conclude that in this case the separation of center of mass and relative motion is shown ~\cite{lol}.

Entering the production and annihilation operators as following \eqref{deq0} and solving Schr\"{o}dinger equation one can derive the known for quantum electrodynamics (QED) solution - the root ambiguity in energy spectrum $\epsilon=\pm\,B\sqrt{l}$, where $l$ is a number of natural numbers set ~\cite{McClure}. The root ambiguity in energy spectrum at the solution of the problem about quantization with relativistic invariance lead in quantum field theory into the creation of a pairs of particles (particles+antiparticles) ~\cite{Berestetskii}. When $l$ is a number of complex numbers set the tachyon solutions are provided by arising the complex energy in spectrum of quantization of Landau for pair of two Majorana fermions coupled via a Coulomb potential.

For graphene with strong Coulomb interaction the Bethe-Salpeter equation for the electron-hole bound state was solved and a tachyonic solution was found ~\cite{Gamayun}.

Calculation model of the graphene reflects continuum symmetry of QED$_{2+1}$ including Lorentz group. SU(2) symmetry are shown to be found similar chiral in the paper ~\cite{Gusynin3}. Hence is conserving quantum number of chirality. In the paper ~\cite{Malard} the selection rules for the electron-radiation and for the electron-phonon interactions at all points in the Brillouin zone are derived based on irreducible representation of the crystallographic space groups as well as the symmetry properties of electrons and phonons. The each of these models are qualitatively different.

In the paper ~\cite{Nandkishore} a chiral superconductivity from electron-electron repulsive in doped graphene in the $M$ point is predicted.

In the paper ~\cite{Rashba} a Magneto-Coulomb levels at a three-dimensional saddle point were found. The Schr\"{o}dinger equation for the three-dimensional saddle surface  geometry at the magnetic field is applied unconventionally was solved exactly in the paper ~\cite{Rashba} by reducing into one-dimensional Schr\"{o}dinger equation.

In the paper ~\cite{Hartmann} the exciton binding energy is scaled with the formed band gap when the magnetic field is applied and an exciton insulator transition in carbon nanotubes was not found and their THz application was predicted.

In the paper ~\cite{Julian} in the UCoGe material the high-temperature superconductivity is connected with spin fluctuations and hence may be reduced by magnetic field is applied.

The exciton Wannier equation for graphene was solved in the papers ~\cite{{lokot},{Stroucken},{Stroucken1},{Malic}}. A very large exciton binding energies were found. In the paper ~\cite{lokot} a theoretical study the both the quantized energies of excitonic states and their wave functions in graphene is presented. An integral two-dimensional Schr\"{o}dinger equation of the electron-hole pairing for a particles with electron-hole symmetry of reflection is exactly solved. The solutions of Schr\"{o}dinger equation in momentum space in graphene by projection the two-dimensional space of momentum on the three-dimensional sphere are found exactly. We analytically solve an integral two-dimensional Schr\"{o}dinger equation of the electron-hole pairing for particles with electron-hole symmetry of reflection. In single-layer graphene (SLG) the electron-hole pairing leads to the exciton insulator states. Quantized spectral series and light absorption rates of the excitonic states which distribute in valence cone are found exactly. If the electron and hole are separated, their energy is higher than if they are paired. The particle-hole symmetry of Dirac equation of layered materials allows perfect pairing between electron Fermi sphere and hole Fermi sphere in the valence cone and conduction cone and hence driving the Cooper instability.

\section{Conclusions}

Schr\"{o}dinger equation for pair of two massless Dirac particles when magnetic field is applied in Landau gauge is solved exactly. Landau quantization $\epsilon=\pm\,B\sqrt{l}$ for pair of two Majorana fermions coupled via a Coulomb potential from massless chiral Dirac equation in cylindric coordinate is found. In this case the separation of center of mass and relative motion is derived. The root ambiguity in energy spectrum leads into Landau quantization for bielectron, when the states in which the one simultaneously exists are allowed. The tachyon solution with imaginary energy in Cooper problem ($\epsilon^{2}<0$) is found. The wave function are shown to be expressed via the associated Laguerre polynomial. In the paper the Cooper problem in superconductor theory is solved as quantum-mechanical problem for two electrons unlike from the paper ~\cite{Gamayun} where the Bethe-Salpeter equation was solved for electron-hole pair. The continuum symmetry of Dirac equation allows perfect pairing between electron Fermi spheres and hence creating a Cooper pair.

\section{Mathematical Appendix}

From a algebraic manipulation one can find a following recurrence relations:

\begin{equation}
\begin{array}{cccc}
G(\alpha,\alpha-\gamma+1,-z)=(F(\alpha,\gamma,z)\frac{\Gamma(\alpha)}{\Gamma(\gamma)}z^{\gamma}-F(\alpha-\gamma+1,2-\gamma,z)\frac{\Gamma(\alpha-\gamma+1)}{\Gamma(2-\gamma)}z)\times\\
\times\frac{\Gamma(1-\alpha)\Gamma(\gamma-\alpha)}{\Gamma(\alpha)\Gamma(1-\alpha)-\Gamma(\gamma-\alpha)\Gamma(\alpha-\gamma+1)(-1)^{\gamma-1}}(-1)^{\alpha}z^{\alpha-\gamma},
\end{array}
\end{equation}
where
\begin{equation}
G(\alpha,\beta,z)=\frac{\Gamma(1-\beta)}{2\pi\,i}\int_{C_{1}}(1+\frac{t}{z})^{-\alpha}t^{\beta-1}e^{t}dt,
\end{equation}

\begin{equation}
F(\alpha,\beta,\gamma,z)=-z^{-\beta}\frac{\Gamma(\gamma)\Gamma(1-\alpha)}{\Gamma(\gamma-\beta)\Gamma(\beta+1-\alpha)}F(\beta,\beta+1-\gamma,\beta+1-\alpha,\frac{1}{z}),
\end{equation}

\begin{equation}
F(\alpha,\beta,\gamma,z)=-z^{-\alpha}\frac{\Gamma(\gamma)\Gamma(1-\beta)}{\Gamma(\gamma-\alpha)\Gamma(\alpha+1-\beta)}F(\alpha,\alpha+1-\gamma,\alpha+1-\beta,\frac{1}{z}),
\end{equation}

\begin{equation}
\begin{array}{cccc}
G(\alpha,\alpha-\gamma+1,-z)=\frac{\Gamma(1-\alpha)\Gamma(\gamma-\alpha)}{\Gamma(\alpha)\Gamma(1-\alpha)-\Gamma(\gamma-\alpha)\Gamma(\alpha-\gamma+1)z^{-2\alpha}}\times\\
\times(\frac{\Gamma(\alpha)}{\Gamma(\gamma)}z^{\gamma-\alpha}F(\alpha,\gamma,z)-\frac{\Gamma(\alpha-\gamma+1)}{\Gamma(2-\gamma)}z^{1-2\alpha}F(\alpha-\gamma+1,2-\gamma,z))e^{-z}z^{2\alpha+\gamma}(-1)^{\alpha},
\end{array}
\end{equation}

\begin{equation}
F(\alpha-\gamma+1,2-\gamma,z)=(-1)^{\alpha-\gamma+1}e^{z}\frac{\Gamma^{2}(\gamma-\alpha)}{\Gamma^{2}(1-\alpha)}F(\alpha,\gamma,-z).
\end{equation}

From a algebraic manipulation one can find a following integrals and recurrence relations which connect theirs:

\begin{equation}
\begin{array}{cccc}
J=\int_{0}^{\infty}e^{-\lambda\,z}z^{\gamma-1}F(-n,\gamma,kz)F(\alpha',\gamma,k'z)dz=\\
\Gamma^{2}(\gamma)\lambda^{\alpha'-\gamma-n}(\lambda-k)^{n}(\lambda-k')^{-\alpha'}\times\\
\times\,(F(\alpha',-n,\gamma,\frac{kk'}{(\lambda-k)(\lambda-k')})-\frac{2k'}{\lambda-k'}F(\alpha'+1,-n,\gamma,\frac{kk'}{(\lambda-k)(\lambda-k')})),\\
\end{array}
\end{equation}

\begin{equation}
\begin{array}{cccc}
J_{\nu}^{s,p}(\alpha,\alpha')=\int_{0}^{\infty}e^{-\frac{k+k'}{2}z}z^{\nu-1+s}F(\alpha,\gamma,kz)F(\alpha',\gamma-p,k'z)dz,\\
\end{array}
\end{equation}

\begin{equation}
\begin{array}{cccc}
J_{\nu}^{s,p}(\alpha,-n)=(-\frac{1}{2\pi\,i})\frac{\Gamma(1-\alpha)\Gamma(\gamma)}{\Gamma(\gamma-\alpha)}\Gamma(\nu+s)\frac{1}{\gamma(\gamma+1)...(\gamma+n-1)}(-1)^{\nu+s-\gamma+p}\frac{n!}{l!(n-l)!}\times\\
\times\,(\gamma-p+n-1)!(\nu+s-\gamma+p)!k'^{l}\lambda^{\gamma}(\lambda-k)^{-\alpha}(\lambda-k')^{-\alpha}\times\\
\times\,(F(\alpha,\alpha,\gamma,\frac{kk'}{(\lambda-k)(\lambda-k')})-\frac{2k}{\lambda-k}F(\alpha+1,\alpha,\gamma,\frac{kk'}{(\lambda-k)(\lambda-k')})),\\
\end{array}
\end{equation}

\begin{equation}
\begin{array}{cccc}
J_{\gamma}^{s,p}(\alpha,\alpha')=\int_{0}^{\infty}e^{-\frac{k+k'}{2}z}z^{\gamma-1+s}F(\alpha,\gamma,kz)F(\alpha',\gamma-p,k'z)dz,\\
\end{array}
\end{equation}

\begin{equation}
\begin{array}{cccc}
J_{\gamma}^{s,p}(\alpha,\alpha')=\frac{\gamma-1}{\gamma-\alpha-1}J_{\gamma-1}^{s+1,p-1}(\alpha,\alpha')-\frac{\alpha}{\gamma-\alpha-1}J_{\gamma}^{s,p}(\alpha+1,\alpha'),\\
\end{array}
\end{equation}

\begin{equation}
\begin{array}{cccc}
(\gamma-\alpha-1)J_{\gamma}^{s,p}(\alpha,\alpha')=(\gamma-1)J_{\gamma-1}^{s+1,p-1}(\alpha,\alpha')-\alpha\,J_{\gamma}^{s,p}(\alpha+1,\alpha'),\\
\end{array}
\end{equation}

\begin{equation}
\begin{array}{cccc}
J_{\gamma}^{s,0}(\alpha,\alpha')=\Gamma(\gamma+s)k'^{-\alpha'}(\frac{k+k'}{2})^{-\gamma-s+\alpha'}\frac{(-1)^{\alpha'-\gamma}(\alpha'-\gamma)!(\gamma+s-1)!}{\gamma(\gamma+1)...(\gamma+s-1)}F(\alpha,\gamma+s-\alpha',\gamma,\frac{2k}{k+k'}).
\end{array}
\end{equation}

\end{widetext}

\end{document}